\documentstyle[psfig]{mn}

\def\wisk#1{\ifmmode{#1}\else{$#1$}\fi}
\def\arcpt{\wisk{''\mkern-7.0mu .\mkern1.4mu}}

\def\ha{H\,$\alpha$}
\def\hb{H\,$\beta$}
\def\hg{H\,$\gamma$}
\def\oiii{[O~{\sc iii}]}
\def\oii{[O~{\sc ii}]}

\def\ciii{C~{\sc iii}]}
\def\civ{C~{\sc iv}}
\def\lya{Ly\,$\alpha$}
\def\feii{Fe~{\sc ii}}
\def\mgii{Mg~{\sc ii}}
\def\nv{N~{\sc v}}

\def\ew{$W_{\lambda}$}

\def\aopt{$\alpha_{\rm opt}$}
\def\arad{$\alpha_{\rm rad}$}
\def\gl{$\lambda$}

\def\gam{$\gamma$}

\newcommand{\degree}{$^{\circ}$}

\title{Origin of the viewing-angle dependence of
the optical continuum emission in quasars}

\author[Joanne C. Baker]
       {Joanne C. Baker\thanks{E-mail: j.c.baker@mrao.cam.ac.uk} \\
  Mullard Radio Astronomy Observatory, 
Cavendish Laboratory, Madingley Road, Cambridge, CB3 0HE, UK. \\
Dept. of Astrophysics and RCfTA, University of Sydney, NSW 2006, 
Australia.  
 }
\date{Accepted 24 Oct 1996}
\pubyear{1996}

\begin{document}
\maketitle


\begin{abstract}

The orientation-dependence of the optical continuum emission in
radio-loud quasars is investigated using a new, complete sample of 
low-frequency-selected quasars, the Molonglo Quasar Sample (MQS).
The optical continuum is found to be highly 
anisotropic, brightening continuously from lobe- to 
core-dominated quasars by 3--5~mag. 
It is argued that aspect-dependent extinction, 
rather than relativistic boosting as has been previously proposed, 
provides the simplest explanation consistent with the data. 
The reddening hypothesis is supported by both the steeper optical 
slopes and the larger Balmer decrements found in lobe-dominated
quasars, as well as the stronger anisotropy seen at blue wavelengths. 
The dust responsible is shown to be physically associated with the
quasar, lying mostly at radii between the broad and narrow-line regions
in a clumpy distribution. Such a geometry is reminiscent of a torus. 
However, substantial numbers of dust clouds must lie within the 
torus opening angle, contributing to an increasing average optical 
depth with increasing viewing angle away from the jet axis, $\theta$.  
In addition, the ratio of \oiii\,\gl4959,5007 to 
\oii\,\gl3727 line flux is shown to be aspect dependent, 
and is consistent with partial obscuration of \oiii\ at large $\theta$.
Trends of broad-line equivalent widths with $R$ are also presented,
including evidence for the luminosity dependence of 
\ew\,(\mgii\,\gl2798).

\end{abstract}



\section{Introduction}
\label{sec:intro}

Attempts to disentangle aspect-dependence from the underlying physics 
of active galactic nuclei (AGN) have been addressed widely under 
the umbrella of the `unified schemes' (see review by Antonucci 1993).
Indeed, orientation has been proposed to be the only difference 
between lobe- and core-dominated 
radio-loud quasars (Orr \& Browne 1982; Kapahi \& Saikia 1982), the latter
viewed from smaller angles to the radio jet and having
Doppler-boosted cores.  

At optical wavelengths the picture is less clear (e.g. Miller 1995). 
Comparisons of core- and lobe-dominated quasars, matched in radio 
flux density, have shown convincingly that core-dominated quasars are 
brighter optically by at least 1~mag 
(Wills \& Lynds 1978; Browne \& Wright 1985). However, 
the precise way in which the optical continuum brightness changes as a 
function of viewing angle is not well constrained: the aim of 
this work is to establish this.  Previous studies 
(Browne \& Murphy 1987; Jackson \& Browne 1989) 
have reported a global decrease of the equivalent 
width of narrow \oiii\,\gl5007 with an increasing ratio, 
$R$, of radio-core to lobe flux density, 
a widely-used orientation indicator 
(Hine \& Scheuer 1980; Orr \& Browne 1982). 
If the narrow \oiii\ emission is assumed to be isotropic, this 
trend points to an increase in optical continuum brightness in 
core-dominated quasars of 2-3~mag (Jackson \& Browne 1989). Although
this large change has been taken as evidence in favour of a 
relativistically-boosted component of optical continuum, the effect is 
weaker than predicted by Doppler-beaming models and would require another 
anisotropic optical component, such as from a disk, to dominate at low $R$.

\begin{figure}
\centerline{\psfig{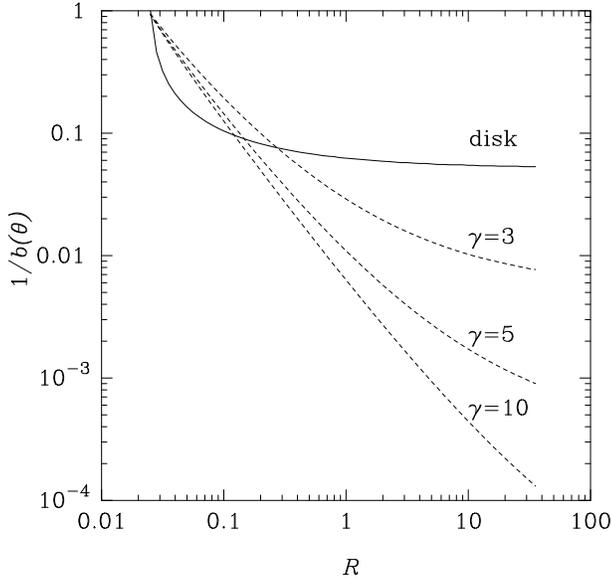} }
\caption[models]
{Simple disk and beaming models for the aspect dependence of the 
optical continuum. Models are shown as the reciprocal of the
continuum enhancement factor, $b(\theta)$, against $R(\theta)$ 
for (i) a thin disk with a limb-darkening factor $a=2$ (Netzer
1987) and (ii) Doppler-boosted synchrotron emission with \gam=3, 5 and 10 
and an intrinsic, unbeamed core-to-lobe ratio 
of $R_{\rm T}=0.024$ (Browne \& Murphy 1987).
   }
\label{bmodels}
\end{figure}

Doppler-boosting will have greatest effect at very small viewing 
angles to the radio jet, i.e. high $R$, reflecting the approximate 
$\gamma^{-2}$ beaming solid angle. The presence of relativistically-boosted 
continuum is associated with the polarised light seen predominantly in 
`blazars' (Impey, Lawrence \& Tapia 1991; Wills et al. 1992).
In contrast, disk emission will show the greatest proportional 
enhancement at low values of $R$ (i.e. with its axis inclined
$\theta\approx 45$\degree\ to the radio jet). 
The variation in optical continuum brightness as a function of $R$ expected 
for a geometrically-thin, optically-thick disk (Netzer 1987) and for
Doppler-boosting models (Orr \& Browne 1982; Browne \& Murphy 1987)
is shown in Figure \ref{bmodels}. 
[These models show the
reciprocal of the continuum enhancement factor, $b(\theta)$, 
for easy comparison with the slopes of later Figures, e.g. 
\ref{rewo3} and \ref{rewo2}].  
Obscuration of the central source at large angles to the jet axis
will also result in anisotropy but with weaker 
angle-dependence than in the beaming case.

A two-component picture has also been proposed at X-ray energies 
(Kembhavi, Feigelson \& Singh 1986; Browne \& Murphy 1987; Baker,
Hunstead \& Brinkmann 1995) where the radio-core and X-ray luminosities
of high-$R$ quasars appear to be tied even more closely. 

Previous optical studies, however, have used data from statistically 
incomplete samples of quasars and are affected by complicated selection
effects. Indeed, as a direct consequence of aspect-dependent optical 
emission, serious orientation biases will affect most 
magnitude-limited samples. For example, Kapahi and Shastri (1987) 
pointed out that the different properties (de Ruiter et al. 1986) of the 
low-frequency-selected 3CR and B2 radio quasar samples could be explained 
fully in terms of selection effects in the fainter B2 sample if the 
optical continuum in quasars were anisotropic.

In this paper, the aspect-dependence of the optical 
continuum is examined for a new, highly-complete  sample of
low-frequency-selected radio quasars, the Molonglo Quasar Sample (MQS).
Radio and optical data have now been obtained for practically the whole 
sample, allowing aspect-dependent trends involving the optical 
continuum, narrow- and broad-line equivalent widths and Balmer 
decrements to be investigated thoroughly. 

\section{The Molonglo Quasar Sample}

\subsection{Selection Criteria}
\label{sec:select}

The Molonglo Quasar Sample (MQS) was drawn initially from the 
408-MHz Molonglo Reference Catalogue (MRC; Large et al. 1981). 
By selecting the sample at low frequency, where the
steep-spectrum extended emission dominates, orientation
biases should be minimised. MRC sources in a $10^{\circ}$ declination strip
($-20^{\circ}>\delta>-30^{\circ}$ and Galactic latitude $|b|>20^{\circ}$)
were selected down to a limiting flux density of 0.95~Jy. `Stellar' optical 
counterparts were sought by eye from the UK Schmidt IIIaJ 
survey plates down to the magnitude limit, $b_{\rm J}\approx22.5$.
Quasar candidates close to the plate limit had their identifications
checked on deep optical $R$-band CCD images ($R \la 25$, taken at 
Las Campanas).

In this way, a total of 113 quasar candidates was found 
(as of 1996 May) from the original MRC source list, resulting in a
quasar detection fraction of about 20\%. 
Spectroscopic confirmation is currently 
available for 108/113 quasars whose redshifts span the range 
$0.1<z<3$. Six BL-Lacs are included in this
number but five of them have no redshift determinations. 
Radio and optical data for the MQS will be published shortly in a
pair of papers: 
Kapahi et al. (Paper I) and Baker et al. (Paper II), in preparation. 
Although the sample size may rise marginally
on completion of the optical identifications for all the radio sources in
the strip, the conclusions of this paper will not be affected
significantly.

\subsection{Radio Observations}

Extensive follow-up radio observations have been undertaken for the MQS, 
including radio imaging with the Molonglo Observatory Synthesis Telescope
(MOST) at 843~MHz and the VLA at 5~GHz. 
The VLA images, with a resolution of about 1\arcsec\ 
and a typical dynamic range of 1000:1 (Paper I), 
have allowed the separation of compact and
extended radio components from which the core-to-lobe flux
density ratios, $R$, have been estimated. 

In addition, the VLA core flux densities of 82 quasars have 
been confirmed independently using the Parkes-Tidbinbilla Interferometer 
(PTI; Norris et al. 1988). The 275-km-baseline PTI was used at 2.3~GHz 
on 1993 Jan 22-25 to probe the radio cores on 0\arcpt 1 scales. The PTI 
flux densities at 2.3~GHz confirmed the 5-GHz VLA measurements for 
strong cores to within 10\%, and typically within 30\% for weak cores 
($<20$~mJy, signal-to-noise ratio $\ga2$), confirming that the radio cores 
have flat spectra and vary little on timescales of several years. 
Following this check, 
VLA 5-GHz core measurements were used in preference for MQS sources with 
clearly defined cores: PTI core flux densities were however used instead, 
under the assumption of a flat core spectrum, if the VLA core was 
blended at 1\arcsec\ resolution. 

Lower limits to $R$ were estimated for core-dominated quasars 
where faint extended structure was swamped by the core emission. 
Conversely, upper limits to $R$ were assigned to several quasars 
where the VLA cores were either blended with steep-spectrum 
components or the radio data precluded easy separation.
The $R$-values have been K-corrected to an emitted frequency of 10~GHz
using spectral indices calculated between 408~MHz and 5~GHz. This 
rest frequency was chosen in order to minimise spectral index 
corrections at the median redshift of the MQS, $z\approx 1$.
$R$-values have not been calculated for the CSS class due to
the complex nature of their radio structures (Fanti \& Fanti 1994), 
and they are considered separately. For this study,  CSS quasars are
defined as having 408~MHz to 5~GHz spectral indices, \arad$ >
0.5$ ($S_{\nu} \propto \nu^{-\alpha}$), 
with no sign of spectral flattening at higher radio frequencies, and 
linear sizes $<20$~kpc. The properties of the CSS quasars will be
addressed in a future paper.

\subsection{Optical Spectroscopy}
\label{sec:aat}

Low-resolution (FWHM 25\,\AA) optical spectra, 
spanning the wavelength range 3400-10000\AA, have been obtained 
for 78  quasar candidates to date with the RGO Spectrograph and FORS on the
Anglo-Australian Telescope (AAT). 
Spectra were taken through a 2\arcsec-wide slit 
oriented at parallactic angle to minimise dispersive losses.
Many spectra were taken under non-photometric conditions, although
typical flux calibration errors lie within a factor of two.
Another two MQS quasars have high signal-to-noise EFOSC 
spectra available (Wall \& Shaver, private communication).
Full observational details and a homogeneous set of 80 spectra were
presented for the MQS in Paper II.  Most spectra have signal-to-noise 
ratios $>10$ in continuum, even for targets as faint as 22~mag. 

Although their classification and redshifts have been confirmed,
spectra for fifteen quasars were unavailable at the time of this analysis
or were judged too poor for emission-line measurements.
Another five candidates await spectroscopic confirmation, 
including two lobe-dominated and three CSS targets.
On the whole, these late identifications were a consequence of ambiguity
due to radio-optical positional offsets and not due to, for example, weak 
emission lines. It is assumed the exclusion of these twenty objects, 
spanning a range of radio properties and optical magnitudes, should 
not introduce significant biases.
As mentioned above, six BL Lacs without emission lines (\ew$<5$\AA) 
are also excluded. One low-redshift ($z= 0.132$) quasar which could be 
otherwise be classified as a broad-line radio galaxy is included, however. 
Another seven quasars have spectral data published in the literature only
and are included in the analysis (see Paper II).

The optical spectra were reduced
using standard techniques within the {\sc starlink figaro} package, as
described in Paper II.  In each case, 
a power-law continuum was fitted by eye and emission-line fluxes 
integrated down to this level. Within a given spectrum, errors in continuum
and line flux are generally large, 20-30\%, due mostly to uncertainties
in positioning the continuum. Smaller errors were measured for the 
emission-line equivalent widths, typically 5-15\%.
Due to the low spectral resolution, broad-line blends, 
such as \lya\,$+$ \nv, have not been deconvolved. 
For the complete sample, all emission-lines in the observed wavelength 
interval were measured (including upper limits) and so the data should 
not be biassed by non-detections.

\section{Statistical Analysis}

Due to the significant numbers of upper and lower limits in the MQS dataset, 
the statistical techniques of survival analysis have been adopted. The ASURV 
package (Rev 1.2; La Valley, Isobe \& Feigelson 1992) 
was used to allow the incorporation of randomly censored datapoints into 
standard correlation and regression tests (Feigelson \& Nelson 1985;
Isobe, Feigelson \& Nelson 1986). Because most tests accept data limits 
in only one direction at a time, lower limits to $R$ only were 
incorporated in most cases as these are most prevalent. Unless 
otherwise stated, Kendall's generalised tau test was used to 
obtain correlation probabilities, $P$(corr), and both the 
EM-Algorithm and Buckley-James methods averaged to obtain 
regression estimates. Where probabilities are stated, all 
the above tests gave similar results, otherwise the coefficients 
were viewed with caution. The probabilities quoted in the following 
sections are given as percentages and indicate the probability that 
the correlation (on the scale plotted) is real.

\section{Anisotropic Optical Continuum Emission}

Direct relationships between the compact radio properties and the
magnitudes and slopes of the optical continuum are explored first.
In common with other studies the ratio $R$ is used as a statistical 
indicator of radio-source orientation (Orr \& Browne 1982)
and (unless otherwise stated) a
typical Lorentz factor $\gamma=5$ is adopted for consistency with 
other work (eg.~Ghisellini, Maraschi \& Treves 1985; Browne \& Murphy 1987). 
Although recent work points to a range of Lorentz factors being present in 
quasar samples (e.g. Padovani \& Urry 1992), in practice this will have
the effect of increasing the dispersion in beamed properties only at
large $R$ (see Figure \ref{bmodels}) and $\gamma=5$ will still be
useful for illustrative purposes.
Values of
$H_{0}=50$~km\,s$^{-1}$Mpc$^{-1}$, $\Omega = 1$ and $\Lambda = 0$
are taken throughout.

\subsection{Optical and Radio-Core Brightness}
\label{sec:mags}

\begin{figure}
\centerline{\psfig{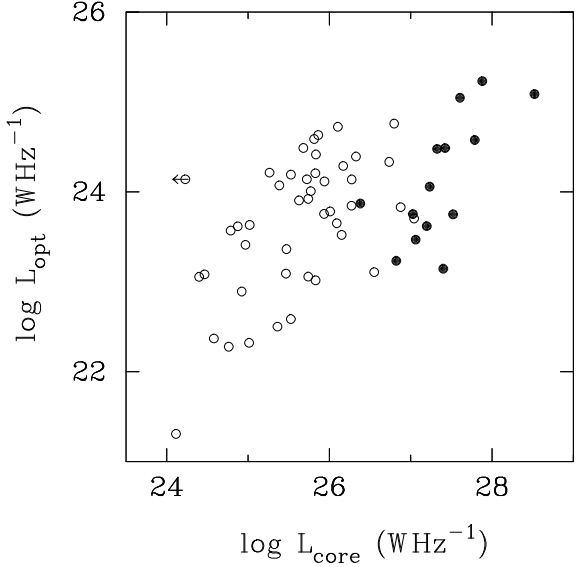} }
\caption[magcore]
{Optical continuum (COSMOS IIIaJ) 
versus 5-GHz radio-core (VLA or PTI) luminosity for MQS quasars.
Core dominated ($R\ge 1$) quasars are plotted as filled circles. CSS sources
have been excluded.
}
\label{magcore}
\end{figure}

The relationship between 5-GHz radio-core and optical luminosity
for the whole sample is plotted in Figure \ref{magcore}. 
The optical luminosity has been calculated  using homogeneous 
optical magnitudes, $b_{\rm J}$, measured from IIIaJ plates 
from the COSMOS Southern Sky Survey (Yentis et al. 1992) at the
Anglo-Australian Observatory. 
Optical spectral indices were measured between 3400 and 10000\AA\ 
from the observed AAT spectra; a mean 
value of \aopt$=0.7$ ($S_{\nu} \propto \nu^{-\alpha}$)
was used otherwise (e.g. Browne \& Murphy 1987). 
Radio-core luminosities assume a flat spectral index.
A trend is visible in Figure \ref{magcore}
such that quasars with luminous radio cores at 5~GHz
are also brighter optically.
The correlation is highly significant ($P>99.99$\%), having an overall
best-fit slope of $0.41 \pm 0.07$. 
Because the MQS is complete, correlated distance effects should be
insignificant, and as confirmation the correlation remains significant
in the flux-flux plane ($P=98$\%).

Core-dominated quasars ($R\ge 1$), highlighted
in Figure \ref{magcore}, lie on a tighter, steeper 
trend than lobe-dominated quasars. The slope, $0.7\pm 0.1$, 
is also slightly steeper than that reported by Browne \& Murphy (1987) 
for a sample of more luminous quasars with $R>1$. 
Although the two luminosities are not proportional, this suggests
a particularly close relationship between the total 
optical and radio-core emission in core-dominated sources, perhaps
due to relativistic beaming. In addition, the correlation 
for core-dominated quasars appears slightly displaced from that for 
lobe-dominated sources. This indicates further 
that the optical brightness is not as strong a function of
viewing angle as the radio, or, that two components of optical emission are 
present in quasars.

\subsection{Optical Magnitudes and $R$}

\begin{figure}
\centerline{\psfig{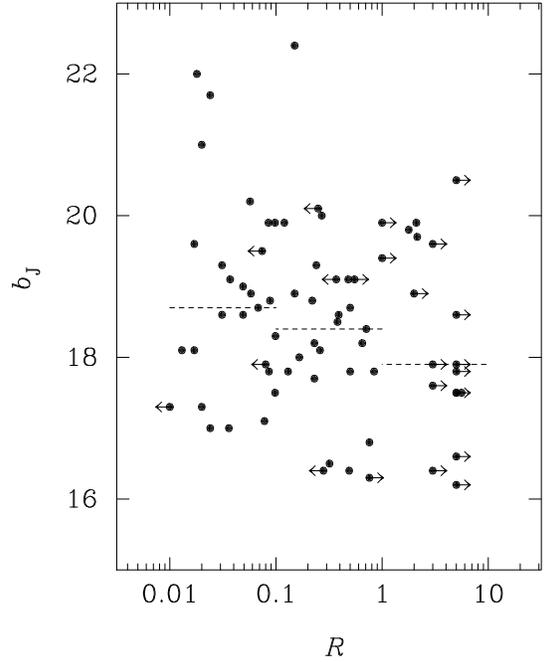} }
\caption[Dist]
{COSMOS IIIaJ optical magnitudes, $b_{\rm J}$, as a function of $R$. 
Limits are marked. Typical errors in $b_{\rm J}$ are $\pm 0.2$~mag.
Median optical magnitudes in three $R$ bins are indicated by dashed lines
($\langle b_{\rm J}\rangle = $18.7, 18.4, 17.9).}
\label{rmag}
\end{figure}

COSMOS optical magnitudes are plotted against $R$ in Figure \ref{rmag}
with group medians superposed. 
A weak trend ($P\approx 90$\%) is observed 
such that core-dominated quasars 
tend to be brighter than their lobe-dominated counterparts by almost 1~mag
in the unbiased MQS.
The median optical magnitudes over three ranges of $R$
support this trend, brightening from $\langle b_{\rm J}\rangle = 18.7$ 
in quasars with $R<0.1$ 
to $\langle b_{\rm J}\rangle = 17.9$ in quasars with $R\ge 1$.
Interestingly, the median optical magnitude of CSS quasars is at least 
1~mag fainter on average ($\langle b_{\rm J}\rangle =19.8$) 
than non-CSS quasars ($\langle b_{\rm J}\rangle =18.6$).

\begin{figure}
\centerline{\psfig{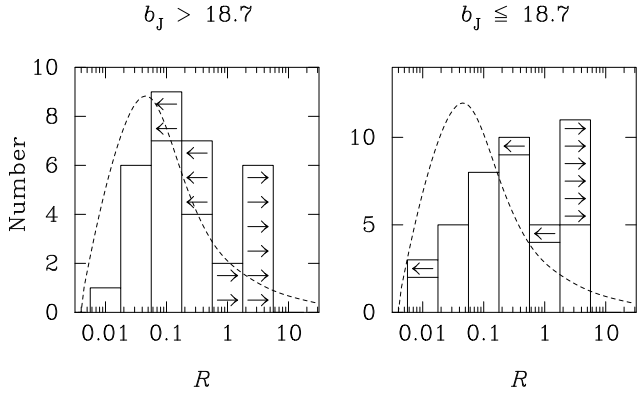} }
\caption[Dist]
{Distribution of radio-core dominance, $R$, for optically-faint and
-bright MQS quasars, split at the median magnitude, $b_{\rm J}=18.7$. 
The expected $R$-distribution for a randomly-oriented sample with
$\gamma=5$ relativistic boosting
is shown as a dashed line for comparison (Orr \& Browne 1982).   
Limits in $R$ are indicated by arrows.
}
\label{rbrfaint}
\end{figure}

Figure \ref{rbrfaint} shows the $R$-distributions for faint and 
bright quasars, separated at the median magnitude of the whole MQS, 
$b_{\rm J}=18.7$ (Paper II). 
The expected $R$-distribution for randomly-oriented quasars, 
according to the relativistic-beaming model of Orr \& Browne (1982) with
$\gamma=5$, is
also shown in Figure \ref{rbrfaint} as a dashed line for comparison. 
Statistical tests confirm that the two $R$-distributions are different 
($P=90$\%). This difference is not a result of intrinsic luminosity 
effects, however, as the redshift distributions of the faint and 
bright sub-samples are indistinguishable by two-sample tests.

Looking at Figure \ref{rbrfaint} more closely, the $R$-distribution 
of faint MQS quasars ($b_{\rm J}>18.7$) appears largely consistent 
with the model distribution for random orientations, except at high $R$. 
In fact, both quasar subsets show an excess of sources with $R>1$, 
suggesting some orientation bias due to the presence of a 
relativistically-beamed component. 
In particular, luminous but more distant quasars will be brought into the
flux-limited sample as a result of extreme Doppler boosting.
Larger numbers of beamed, distant quasars would be expected in
the sample if the distribution of Lorentz factors had a tail
to high values $\gamma \gg 5$.
This could explain the proportion of core-dominated 
quasars in the MQS of about 25\%, somewhat higher than the 20\% predicted 
by the simple model of Orr \& Browne with a single $\gamma \approx 5$.
Also, the $R$-distribution for the brighter 
($b_{\rm J} \le 18.7$) sub-sample in Figure \ref{rbrfaint} 
peaks at higher $R$-values than for fainter quasars
by about a factor of ten (Kapahi, Subrahmanya \& D'Silva 1989). 
This indicates that brighter quasars tend also to have stronger cores, 
but they are not necessarily the most highly boosted. Such a picture 
suggests that anisotropic processes occurring at lower $R$ are 
dominated by a process other than relativistic boosting.

\begin{figure}
\centerline{\psfig{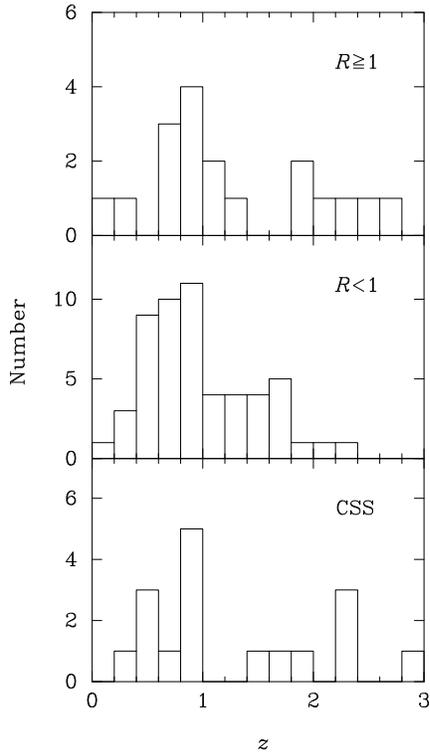} }
\caption[Dist]
{Redshift distributions for MQS quasars with $R\ge 1$, $R<1$ and CSS
radio morphologies.
}
\label{rzdis}
\end{figure}

The differences in Figure \ref{rbrfaint} become stronger if the sample
is split in optical luminosity rather than apparent magnitude. 
Redshift-dependent selection effects should have little effect on
luminosity trends for the MQS as the distributions of redshift 
are similar for quasars with $R \ge 1$ and $R<1$.
Figure \ref{rzdis} shows the distributions of redshift for 
MQS quasars with $R \ge 1$, $R<1$ and CSS radio morphologies.
For the three subsets, the median redshifts are 0.99, 0.81 and 0.78,
respectively, and they span similar ranges.

\subsection{Optical Spectral Slope and $R$}
\label{sec:raopt}
 
\begin{figure}
\centerline{\psfig{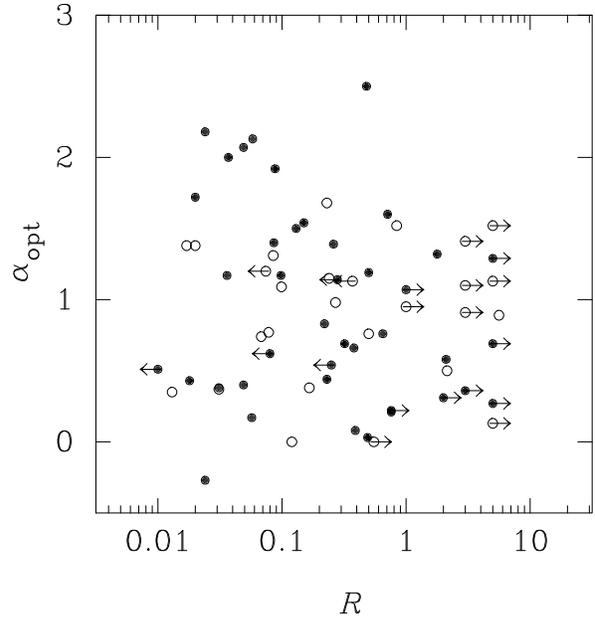} }
\caption[$R$-Dependence of aopt]
{$R$-dependence of \aopt\ (3400--10000\AA) for the MQS. 
High-redshift ($z\ge 1$) quasars are marked by open circles, 
low-redshift ($z<1$) ones by filled symbols. Errors in \aopt\ 
are 20--30\%. The highest-\aopt\ quasar on the plot, 2024--217, 
also has a large Balmer decrement (Paper II), 
suggesting it is reddened unusually highly.
}
\label{raopt}
\end{figure}

Together with the brightening of the optical continuum in core-dominated 
quasars, as inferred from Figures \ref{magcore} and \ref{rbrfaint}, 
any observed change of continuum slope might give a clue to 
the mechanism for the enhancement. Figure \ref{raopt}
shows the observed (3400--10000\,\AA) optical spectral index, \aopt,
plotted as a function of $R$. 
High- and low-redshift MQS quasars have been marked
separately because numerous line blends, particularly the 3000-\AA\ bump
(Wills, Netzer \& Wills 1985), 
make it more likely that the true continuum is masked at the restframe
wavelengths observed in the high-redshift objects (see Paper II). 

Although the scatter in Figure \ref{raopt} is very large, an upper 
envelope is apparent in the low-$z$ subset, 
such that the maximum observed \aopt\ decreases towards high $R$. 
At least, the reddest quasars are all at low $R$.
In this low-redshift subset  the 
continuum slope measurement is more reliable.  As mentioned in 
Section \ref{sec:aat}, a number of $z>1$ quasars 
have been left out of this plot awaiting good quality optical spectra
(see Paper II), and their final inclusion will test whether the apparent
paucity of red quasars at high redshifts is real, or a temporary 
selection effect. The spectral properties of quasars as a function of 
\aopt\ is discussed later (see Section \ref{sec:discussion}).

As shown in Paper II, a correlation between optical magnitude and spectral
index also exists in the MQS, such that flatter optical continua are 
found preferentially in optically-bright quasars. 
The relationships of \aopt\ with both $b_{\rm J}$ 
and $R$, along with the dependence of optical magnitude on $R$ (Figure 
\ref{rbrfaint}), suggest that the optical continuum is both 
brighter and harder in core-dominated quasars. 
As selection effects work in the opposite direction, 
the simplest explanation is that bright, red quasars are rarer 
than bright, blue ones because they are drawn from a higher 
luminosity population. 
 
The large range of optical spectral indices observed in the MQS
could be interpreted in terms of
two components of optical continuum emission in 
quasars --- a hard component which dominates at small angles 
and a steep-spectrum component which is strong at larger 
inclinations. Or, the greater tendency for lobe-dominated 
quasars to have red spectra could point to aspect-dependent 
dust extinction as the culprit.

\section{Narrow-Line Equivalent Widths}
\label{sec:nlews}

In order to correct for intrinsic luminosity effects and therefore
reveal the aspect-dependence of the optical continuum in a flux-limited
sample, an orientationally-independent property is needed as a benchmark.
Extended narrow-line emission is a natural choice as it is mostly emitted 
far ($>1$\,kpc) from the nucleus and is thought to be 
isotropic (Pierce \& Stockton 1986; 
Stockton \& MacKenty 1987; Baum \& Heckman 1989). 
Previous studies (Wills \& Browne 1986; Browne \& Murphy 1987;
Jackson \& Browne 1989) tended to use the narrow 
\oiii\,\gl5007 line for this comparison because it is prominent and occurs 
in a spectral region relatively free from line blends. 
 
Equivalent widths, \ew, have been measured for the MQS for a range of strong 
emission lines (Paper II). The $R$-dependence 
of the equivalent widths is now investigated, beginning with the narrow lines.
CSS sources are excluded from correlations 
involving $R$, but are otherwise included.

\subsection{Equivalent Widths of \oiii}

\begin{figure}
\centerline{\psfig{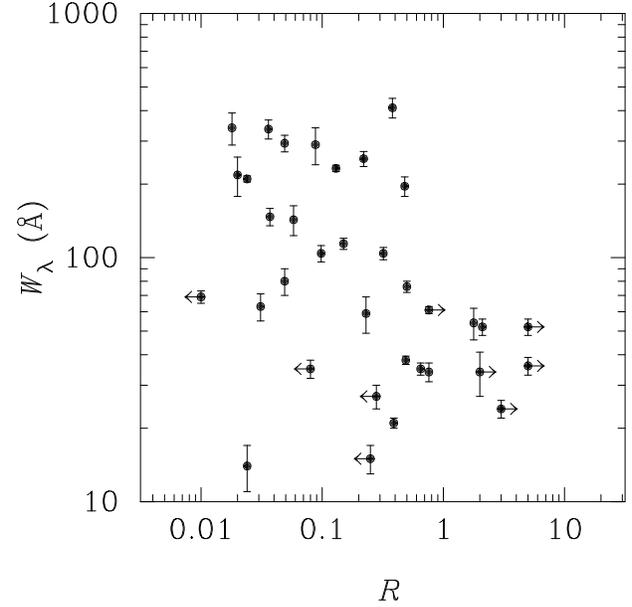} }
\caption[EW of \oiii\ with $R$]
{$R$-dependence of the equivalent width, \ew, of narrow
\oiii\gl\gl4959,5007.  
Error bars indicate the measurement errors for each object. Lower and 
upper limits to $R$ are shown appropriately with arrows. 
   }
\label{rewo3}
\end{figure}

Figure \ref{rewo3} shows the \ew$-R$ relationship for
the \oiii\,\gl\gl4959,5007 doublet for the MQS, 
limited observationally to quasars with $z < 0.9$. 
Although the scatter on Figure \ref{rewo3} is substantial,
a clear anticorrelation is apparent with a Kendall's tau likelihood 
of being real of 99.7\%, upon inclusion of the lower limits to $R$.
The anticorrelation in Figure \ref{rewo3} spans about a factor of ten in
\ew, in quantitative agreement with the earlier data of Jackson 
\& Browne (1989). Again, this is less than expected by Doppler-boosting
alone (see Figure \ref{bmodels}) but is consistent with two-component models
for the optical continuum. 
The large scatter on the plot cannot be attributed solely to  
measurement errors; factors contributing to the scatter are discussed later.

\subsection{Equivalent Widths of \oii}
 
\begin{figure}
\centerline{\psfig{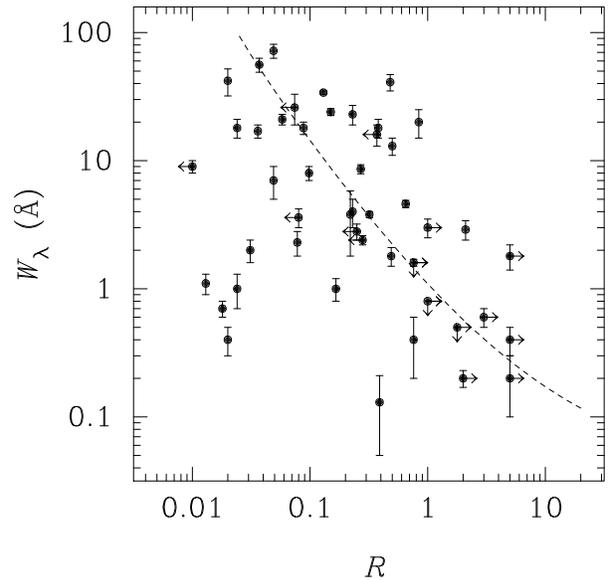} }
\caption[Equivalent Width of \oii\ with $R$]
{$R$-dependence of the equivalent width, \ew, of narrow
\oii\,\gl3727. 
As a guide, the dashed line shows the slope of the variation 
expected from a $\gamma=5$ beaming model (see text) --- 
note the line has an arbitrary vertical offset on a log scale, depending
on the assumed intrinsic unbeamed value of \ew.
   }
\label{rewo2}
\end{figure}

Due to the high signal-to-noise ratio and blue coverage of the MQS spectra, 
narrow \oii\,\gl3727 equivalent widths have also been measured for 
a large number of quasars (up to $z\approx 1.5$). 
In Figure \ref{rewo2}, \ew\,(\oii) is plotted against
$R$ for MQS quasars (note the scale is different from Figure \ref{rewo3}). 
Despite the large scatter, a significant anticorrelation is observed
($P= 99.7$\%), with  core-dominated 
quasars having smaller \ew\,(\oii) than their 
lobe-dominated counterparts by factors of 100--1000. 
Such a decrease is remarkably consistent with the expected 
aspect-dependence of a relativistically-beamed optical continuum 
component.

Measurement uncertainties are larger, on average, for
weaker \oii\ than \oiii\ due to the underlying broad 3000-\AA\ restframe
feature. To minimise this uncertainty, the polynomial continuum fit was tied 
when possible to the \hb--\oiii\ region and extrapolated out to 
the wavelength of \oii. The \ew-measurement errors shown take 
account of the difference 
between using the extrapolated or local continuum level.
This approach is justified {\it a posteriori\/} by the
finding that the 3000\AA\ bump strength appears to depend on $R$
(Baker \& Hunstead 1995). However, in quasars with $0.9<z<1.5$, little 
line-free continuum is available and the equivalent width may be
underestimated by up to 50\% in individual cases, e.g.
several high-$z$, low-\ew, low-$R$ objects in Figure \ref{rewo2}. 
Over the range available, 
no redshift dependence was found for \ew\,(\oii), ruling out any 
strong luminosity dependence affecting higher-$z$ quasars.
 
The strong decrease of \ew\,(\oii) with $R$ in the MQS indicates that the
continuum is highly anisotropic. However, the weaker trend observed for \oiii\
implies that the line luminosities themselves, or their ratio, must 
be $R$-dependent as a consequence of obscuration or ionisation effects.

\subsection{$R$-Dependence of the \oii/\oiii\ Ratio}
\label{sec:roratio}
 
\begin{figure}
\centerline{\psfig{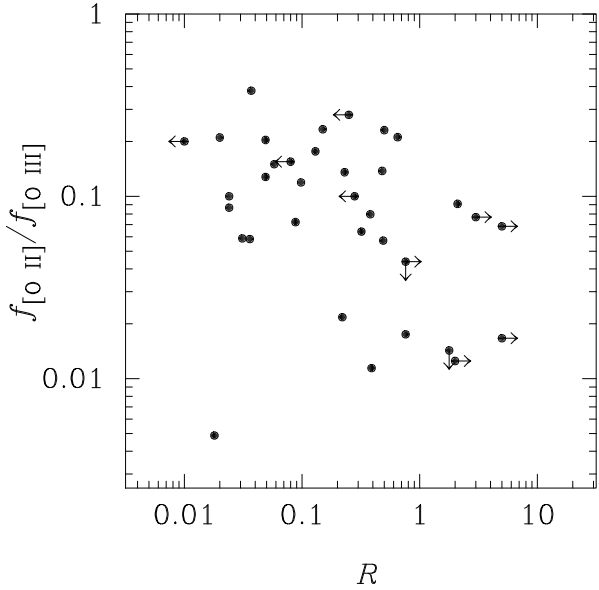} }
\caption[$R$-Dependence of \oii/\oiii\ Ratio]
{$R$-dependence of the \oii\gl3727/\oiii\gl\gl4959,5007 flux ratio for 
MQS quasars. Errors in 
relative line flux are $\approx 20$\%.
 }
\label{ro2o3}
\end{figure}

Figure \ref{ro2o3} shows the ratio of \oii\,\gl3727/\oiii\,\gl4959$+$5007
line flux as a function of $R$ for the MQS. 
A strong anticorrelation is found ($P=98$\%),
the ratio \oii/\oiii\ falling off by at least a factor of ten. 
(Note that one quasar stands out on the plot, 1247-290, 
having both low $R$ and \oii/\oiii\ --- its spectrum in Paper II 
confirms that \oii\ is unusually weak in this object.)
This trend might imply that either
\oii\ or \oiii\ emission is anisotropic. 
Alternatively, high-$R$ quasars on the plot may produce harder 
ionising radiation, thus reducing the \oii/\oiii\ ratio, for example if
they have more powerful jets (see Section \ref{sec:discussion}).

\section{Broad-Line Equivalent Widths}

Originating closer to the quasar nucleus, perhaps in a disk or wind, 
broad-line emission is likely to be anisotropic 
(Osterbrock \& Mathews 1986; Collin-Souffrin 1992; Peterson et al. 1992), 
and a relationship with $R$ is of additional interest in trying to 
understand the broad-line region (BLR).

\begin{figure}
a)
\centerline{\psfig{file=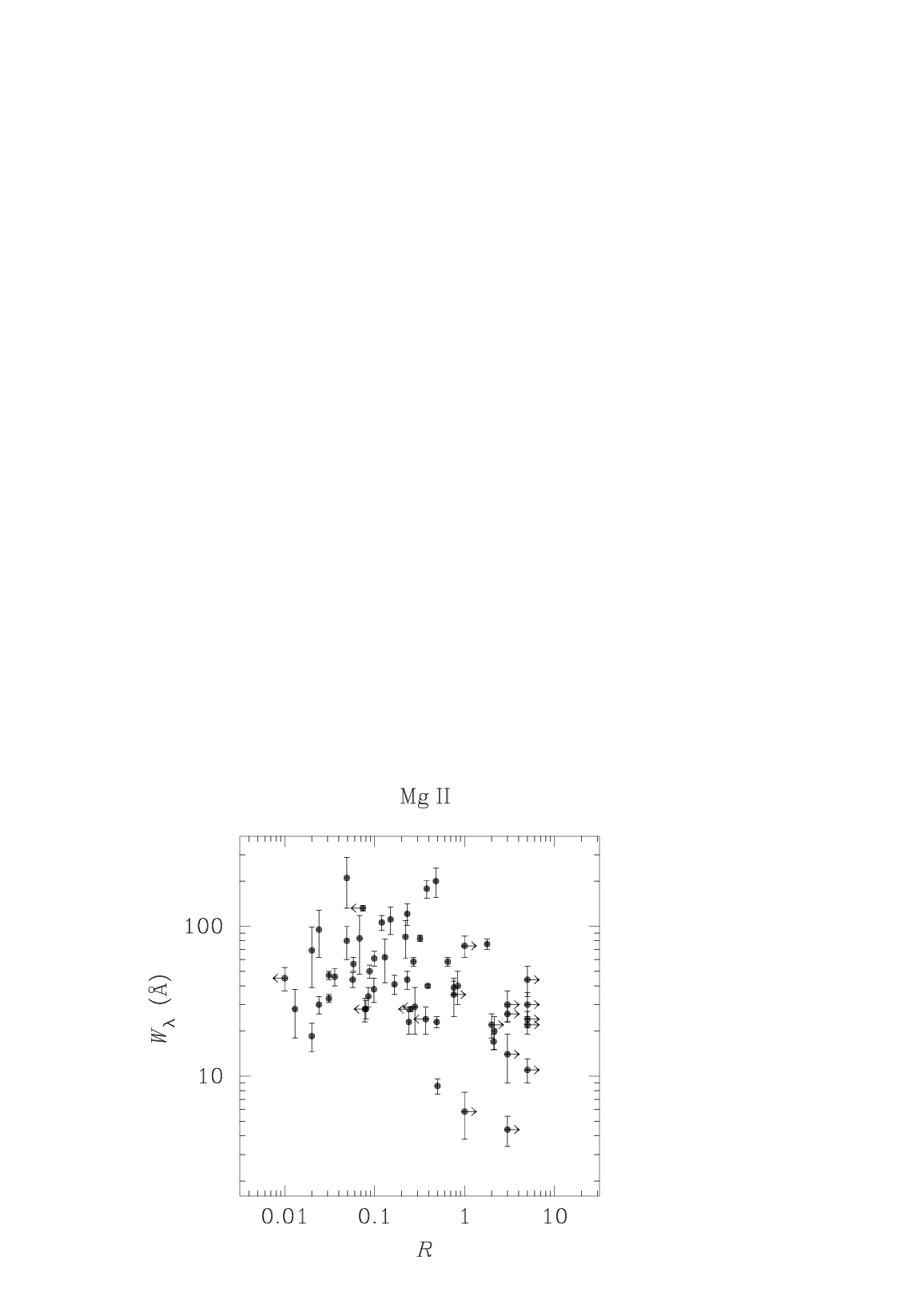,bbllx=84pt,bblly=35pt,bburx=
365pt,bbury=325pt,clip=,height=7cm} }
b)
\centerline{\psfig{file=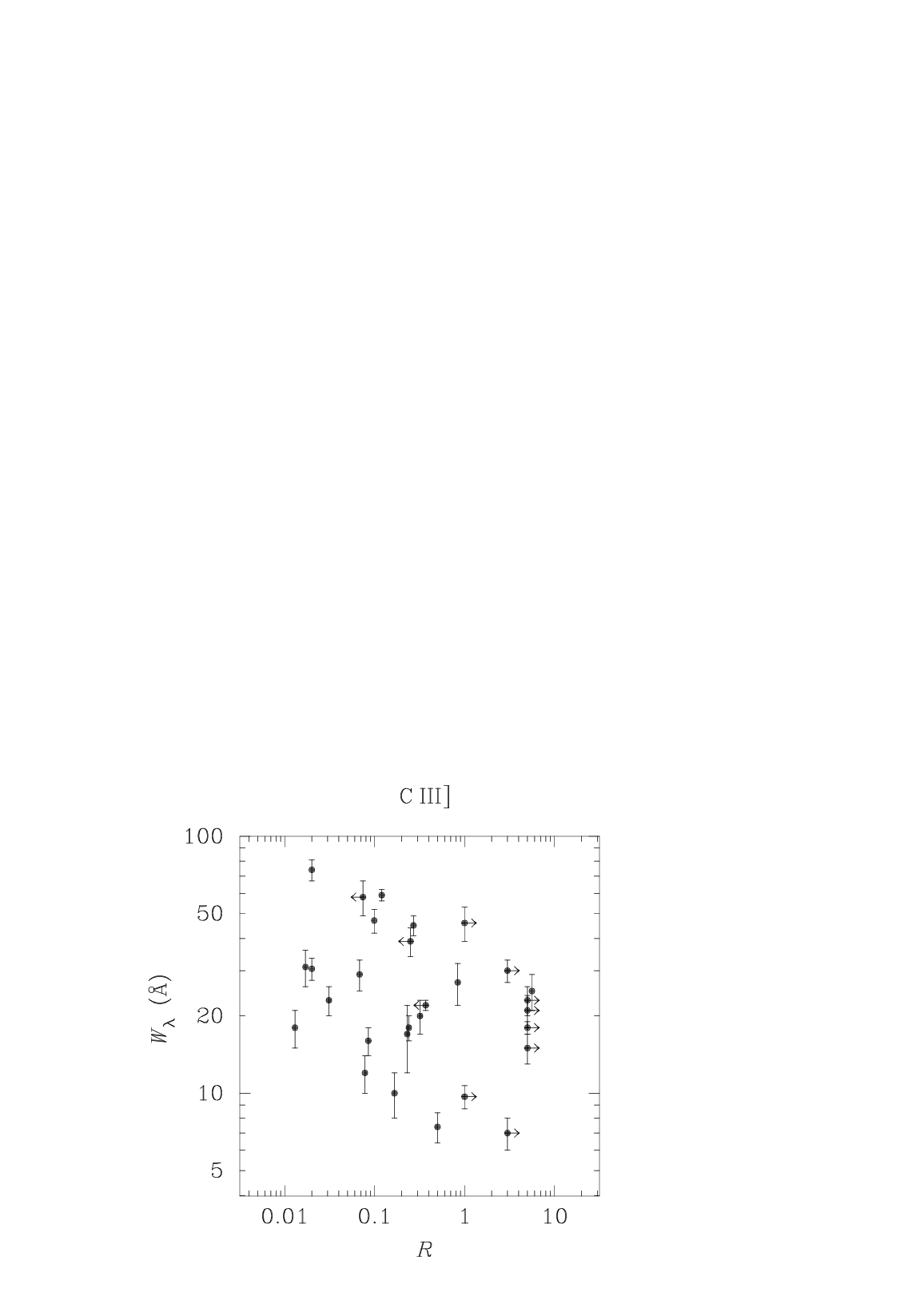,bbllx=84pt,bblly=35pt,bburx=
365pt,bbury=325pt,clip=,height=7cm} }
c)
\centerline{\psfig{file=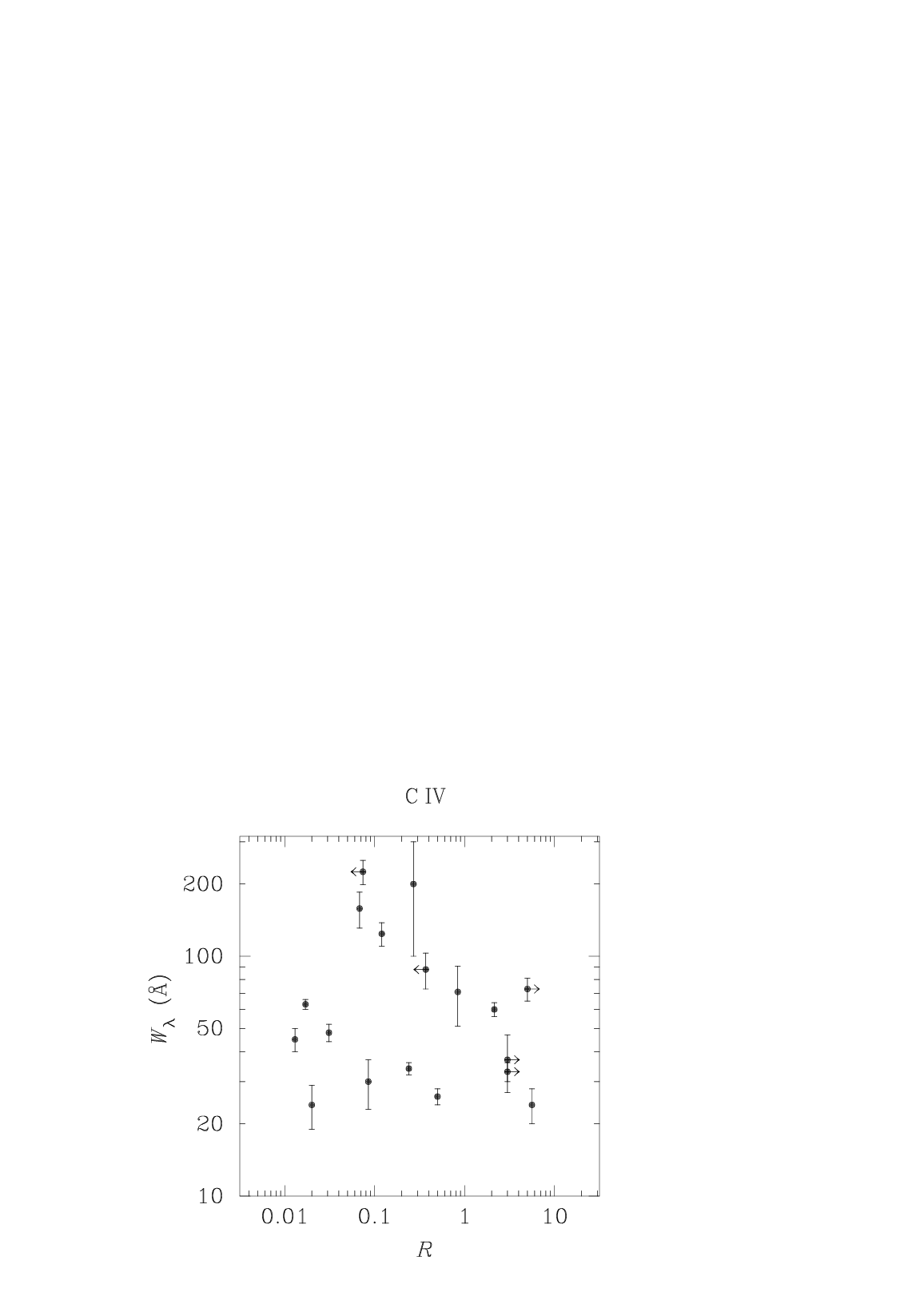,bbllx=84pt,bblly=35pt,bburx=
365pt,bbury=325pt,clip=,height=7cm} }
\caption[$R$-Dependence of \mgii, \ciii\ and \civ\  Equivalent Widths]
{$R$-dependence of the equivalent width, \ew, of a) \mgii\,\gl2798,
b) \ciii\,\gl1909 and c) \civ\,\gl1549, for MQS quasars.
 }
\label{rewbls}
\end{figure}

\subsection{Broad \mgii, \ciii\ and \civ.}
 
The $R$-dependence of the equivalent widths of the
broad lines \mgii\,\gl2798,  \ciii\,\gl1909 
and \civ\,\gl1549  is presented in Figure \ref{rewbls}.
A strong dependence of \ew\,(\mgii) on $R$ is present 
in Figure \ref{rewbls}a) with 99.7\% significance in the unbiased MQS. 
\mgii\,\gl2798 is the most commonly-detected line over the 
redshift range of the MQS; data have been measured for 67 quasars 
with $z<2.2$.  A similar anticorrelation of \ew\,(\mgii) with $R$
was reported by Browne \& Murphy (1987) and 
Baldwin, Wampler \& Gaskell (1989) although 
these samples are affected more strongly by selection effects.

No significant trends with $R$ are found for \ew\
\ciii\ and \civ, although these plots include less points than for \mgii.
Both plots are, however, qualitatively consistent with the 
\mgii\ trend, and show a paucity of high-\ew\ 
core-dominated quasars, i.e. the upper right-hand
corner is empty. Such bright, strong-lined objects  should not have been
missed due to the selection criteria.

\begin{figure}
\centerline{\psfig{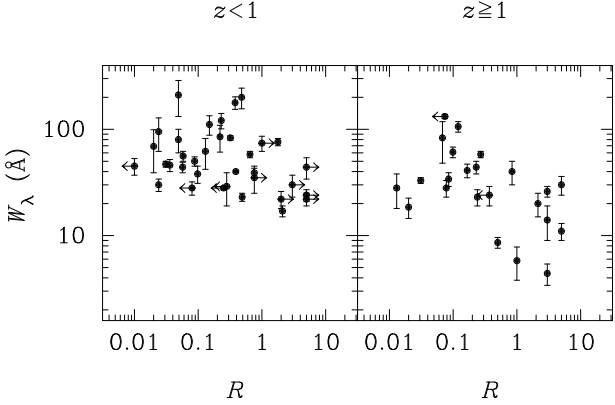} }
\caption[\ew$-R$ for \mgii\ Separated in Redshift]
{$R$-dependence of the equivalent width, \ew, of \mgii\,\gl2798
for high and low redshift MQS quasars, separately. 
}
\label{rewmsep}
\end{figure}
 
On closer inspection, the \mgii\ anticorrelation in Figure \ref{rewbls}
shows a strong redshift dependence, having a much steeper gradient
in the $z>1$ subsample (Figure \ref{rewmsep}). Both subsamples are well
populated with data and show significant anticorrelations. 
The stronger $R$-dependence at high-$z$ cannot be explained solely by
overestimation of the observed continuum level 
due to the presence of the 3000-\AA\ bump, as this is expected to 
vary by less than a factor of two in \ew\ (see Baker \& Hunstead 1995).
Therefore, this trend must reflect a real luminosity dependence or evolution 
of the quasar spectra.

\subsection{The Balmer Lines}

\begin{figure}
\centerline{\psfig{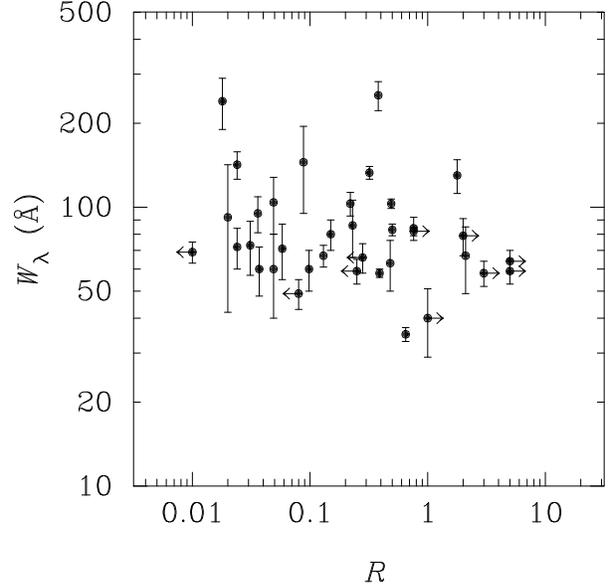} }
\caption[Equivalent Width of \hb\ with $R$]
{$R$-dependence of \hb\ equivalent width, \ew, for MQS quasars.
   }
\label{rewhb}
\end{figure}

Figure \ref{rewhb} shows the $R$-dependence of the equivalent widths of 
\hb\ for MQS quasars with $z<1$.  No strong dependence of \ew\,(\hb) 
is observed, although the plot is qualitatively consistent with
those in Figure \ref{rewbls}. 
The small dispersion, roughly a factor of two, agrees with
the previously-reported close proportionality between the non-thermal 
optical-continuum luminosity and \hb-line luminosity seen in all types
of AGN (Yee 1980; Shuder 1981) and taken to imply a direct link via 
photoionisation. This confirms that the relationship is not an artifact
due to selection effects in the early samples.

\begin{figure}
\centerline{\psfig{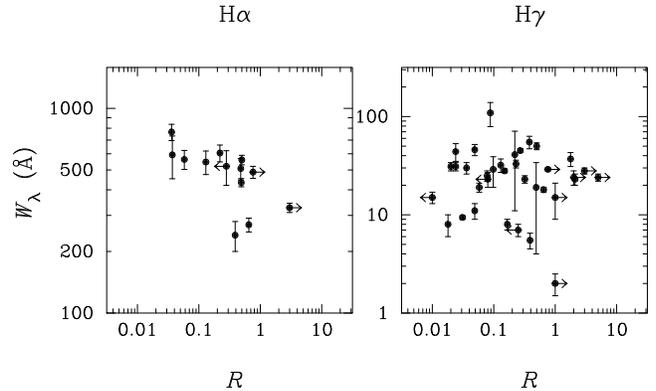} }
\caption[Equivalent Widths of \ha\ and \hg\ with $R$]
{$R$-dependence of \ha\ ({\it left\/}) and \hg\
({\it right\/}) 
equivalent widths, \ew, for MQS quasars with $z<0.5$ and $z<1.2$ 
respectively.
   }
\label{rewhahg}
\end{figure}

Figure \ref{rewhahg} presents the \ew$-R$ plots for the \ha\,\gl6563 and 
\hg\,\gl4340 lines.  \ew(\hg) shows the greater scatter in Figure 
\ref{rewhahg}, probably indicating substantial contamination by 
\oiii\,\gl4363.
Although the number of quasars with
measurable \ha\ ($z<0.5$) is small, there is a formal correlation 
probability of 99.5\% that \ew\,(\ha) decreases at high $R$.
\ew\,(\hg) shows no significant anticorrelation with $R$. 
Again, both plots are still consistent with the trends 
in Figure \ref{rewbls}, notably the paucity of points in the
high-\ew, high-$R$ quadrant.
  
\subsection{Balmer Decrements}
\label{sec:rbalmer}
 
\begin{figure}
\centerline{\psfig{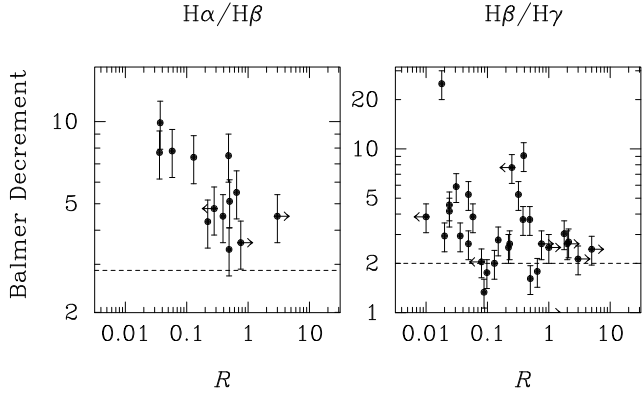} }
\caption[$R$-Dependence of Balmer Decrements]
{$R$-dependence of Balmer decrements for MQS quasars, i.e. 
the ratios \ha/\hb\ ({\it left\/}) and  \hb/\hg\ ({\it right\/})
depending on redshift. 
Expected (case-B recombination) values are shown as dashed lines.
The anticorrelation remains significant even if the quasar with 
\hb/\hg\,$>20$ (MQS\,1247$-$290) is removed.
   }
\label{baldecr}
\end{figure}

The Balmer-line ratios \ha/\hb\ and \hb/\hg\ are now investigated.
Figure \ref{baldecr} shows that both Balmer decrements 
tend to decrease with $R$.
The anticorrelation for \ha/\hb\ has a 97\% significance despite including
only 13 points. The \hb/\hg\ correlation is weaker than
\ha/\hb\ ($P$(corr)$> 90$\% excluding the extreme point) and also
shows more scatter, probably due to significant 
contamination of \hg\ by \oiii\,\gl4363. A similar decrease in 
\hb/\hg\ ratio with $R$ was reported previously by Jackson \& Browne (1991)
for a heterogeneous sample.
Blending of \ha\ and \hb\ with narrow lines or \feii\  will
contribute to the scatter. 

The Balmer decrements of core-dominated quasars ($R\ge 1$)
approach the canonical `case-B' \ha/\hb\ value of 2.85 
(e.g.~Davidson \& Netzer 1979; Osterbrock 1989), suggesting that these
simple assumptions are valid for these quasars.
In contrast, quasars at low $R$ show \ha/\hb\ ratios of up 
to ten, implying they are reddened.
The $R$-dependence of \ha/\hb\ suggests that  quasars viewed 
at large angles to their radio-jet axis suffer greater extinction than
those viewed pole on.

\section{Discussion}
\label{sec:discussion}

The correlations and trends presented above confirm that
core-dominated MQS quasars are indeed brighter optically than their
lobe-dominated counterparts. This is clear from direct
comparison of optical magnitudes in the unbiased MQS  
(Figures \ref{rmag} and \ref{rbrfaint}) 
and is even more convincing if the optical continuum is 
normalised to more isotropic components, such as the narrow-line 
flux (Figures \ref{rewo3} and \ref{rewo2}). In addition, Figures
\ref{raopt} and \ref{baldecr} show that lobe-dominated quasars have both 
redder continua and larger \ha/\hb\ ratios than core-dominated 
quasars. These results are now discussed and a consistent picture proposed.

\subsection{Narrow-Line Equivalent Widths}
\label{sec:discnlews}
 
The decrease with increasing $R$ of both the narrow \oiii\ and \oii\ 
equivalent widths (Figures \ref{rewo3} and \ref{rewo2}) suggests the optical 
continuum is brighter in core-dominated quasars by a factor of
between 10 and 100. However, the amount by which the \ew\ of the 
two narrow lines drops is different. If the drop were due entirely to
continuum changes, this would imply the continuum strength 
changes more in the blue (see Figure \ref{nlewaopt} and later discussion).
Otherwise,  either \oii\ or \oiii\ must be emitted anisotropically, 
or their ratio must change with $R$. 
In support of \oiii\ being anisotropic, \oiii\ has been shown to be weaker 
in radio galaxies than quasars of matched extended 
luminosity (Jackson \& Browne 1990; Jackson \& Eracleous 1995). 
It has been argued that this is due to obscuration of the 
nuclear \oiii\ component in radio galaxies. 
In contrast, the indistinguishability of \oii\ line luminosity 
between radio galaxies and quasars matched 
in extended radio flux density (Bremer et al. 1992; Hes, 
Barthel \& Fosbury 1993) suggests \oii\ emission is more isotropic.
As is clear in Figure \ref{ro2o3}, the \oii/\oiii\ line ratio itself
is $R$- dependent, consistent with partial obscuration of \oiii\ in 
lobe-dominated quasars.

Alternatively, the overall decrease in \ew\ at high $R$ may arise if
core-dominated quasars in the sample include highly-beamed objects of low 
instrinsic power, with weak extended radio and
narrow-line emission.  This argument follows from the
global correlation observed between narrow-line and radio luminosity for
radio-loud AGN (Saunders et al. 1989; Rawlings 1994). However, 
the majority of core-dominated quasars in the MQS still have 
low-frequency flux densities well in excess of 
the measured core flux (confirmed by the 0\arcpt 1 resolution PTI 
flux densities)  and would remain in the sample on the basis of their 
extended emission alone. No significant trends were found between 
the radio and narrow-line luminosities of MQS quasars, as expected
over the relatively small radio luminosity range of the 
MQS ($\sim$1--2 decades). This makes it unikely that
intrinsic power effects dominate the \ew -$R$ plots. 
However, the strength of
\oii\ relative to \oiii\ may be reduced (by about an order of magnitude) 
in some luminous core-dominated quasars as a consequence of an $R$- or 
luminosity-dependent ionisation parameter (e.g. Saunders et al. 1989; 
Netzer 1990).

At low $R$, the small numbers of objects with very high \ew\ suggests 
that most of these are excluded from quasar samples 
(e.g. Francis 1993; Zwitter \& Calvani 
1994), for example if they were classified as radio galaxies. Spectroscopy of
a complete sample of 2-Jy southern radio sources by Morganti et al. (1995)
shows that radio galaxies do indeed have larger \oiii\ and \oii\
equivalent widths than comparable quasars. These are also consistent
with the trends shown for the MQS.

\subsection{Scatter on \ew$-R$ Plots}
\label{sec:scatter}
 
The scatter on the \ew$-R$ plots for both \oii\ and \oiii\ 
(Figures \ref{rewo3} and \ref{rewo2}) is very large. It is larger than
the measurement errors, and so a number of intrinsic processes presumably
contribute.  

Detailed investigation of the effects of radio and optical 
variability is precluded by the present single-epoch data, 
although variability is expected to be small for lobe-dominated quasars
in any case. Radio-core variations (from comparing VLA and PTI data, 
assuming a flat spectral index) were less than a factor of two for 
all sources. Optical continuum variability 
in quasars has been estimated to contribute only 0.2 magnitudes 
on average (Treves et al. 1989; Hook et al. 1991), and
line variability is negligable within the measurement uncertainties,
especially for the narrow lines. The 
effect of missing extended line flux through a narrow slit
(see Paper II) is also negligable over the sample.

Extended emission-line regions are ubiquitous in AGN and the total energy
output in narrow-line luminosity is probably related closely to the power
of the radio jet emanating from the nucleus.
The observed scaling of the narrow-line luminosity with 
radio-jet power (Saunders et al. 1989) itself shows scatter 
comparable with that seen in Figures \ref{rewo3} and \ref{rewo2}.  
In addition, the intensity of \oii\ is known to be more sensitive 
than \oiii\ to the ionisation parameter (Netzer 1990) and intrinsic 
differences in ionisation level will also cause a spread in the 
ratio of \oii/\oiii\  (Figure \ref{ro2o3}). Differences in local
environment may also cause dispersion in narrow-line luminosities
(e.g. McCarthy 1993). Scatter in \ew\ may also increase due to
variations in the broad-band continuum shape itself 
from source to source (Elvis 1994).

\subsection{Reddening in MQS Spectra}

\begin{figure}
\centerline{\psfig{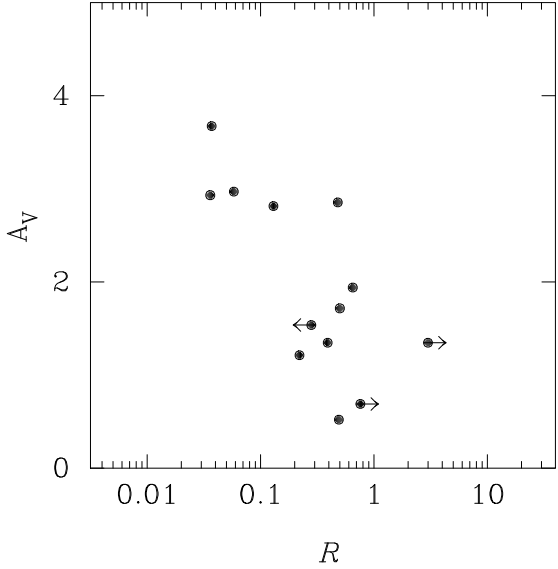} }
\caption[Avr]
{Visual extinction, $A_{\rm V}$ derived from the \ha/\hb\ ratio, 
as a function of $R$ for the MQS. Typical errors are 30\%.
  }
\label{avr}
\end{figure}

 The combination of trends with $R$, especially
the decrease in  \ha/\hb\ (Figure \ref{baldecr}) 
and \aopt\ (Figure \ref{raopt}) and the changing  relationship between the 
radio-core and optical luminosity (Figure \ref{magcore}) with increasing
$R$, suggests that the continuum anistropy is not due solely to
relativistic beaming. Extinction would seem to
provide the explanation: this is investigated further below.
To illustrate the increasing extinction at low $R$,
Figure \ref{avr} plots the implied visual extinction, $A_{\rm V}$, 
against $R$. The estimated  values of $A_{\rm V}$ were converted 
from the \ha/\hb\ Balmer decrements (Figure \ref{baldecr})
assuming a roughly $\lambda^{-1}$ reddening law (e.g.~Osterbrock 1989).

\begin{figure}
\centerline{\psfig{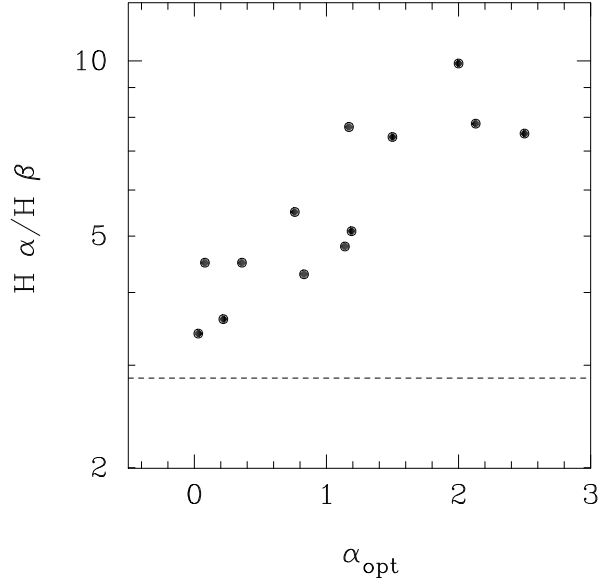} }
\caption[Dependence of Balmer Decrements on Optical Spectral Index]
{Dependence of \ha/\hb\
on optical spectral index, \aopt, for the MQS. 
CSS quasars are excluded.
The canonical value  \ha/\hb$=2.85$ is shown as a dashed line 
(Osterbrock 1989). 
  }
\label{balaopt}
\end{figure}

Figure \ref{balaopt} shows the remarkably tight correlation ($P=99.9$\%,
slope $0.16\pm 0.03$)) between \ha/\hb\ and 
\aopt\ for low-redshift MQS quasars. This confirms that the two 
are closely associated, presumably a consequence of dust extinction.
This clear association supports the use of both \ha/\hb\ and \aopt\ 
as reddening indicators, with additional caveats. We assume \aopt\ is
still representative once \ha\ has been redshifted beyond $\sim9000$\AA. 

It is noted that broad Balmer-line ratios may be 
affected by collisional excitation and optical-depth 
effects within the BLR (e.g. MacAlpine 1985) and this will 
contribute to the scatter. \ha/\hb\
ratios up to $10$ can be reproduced in such models. However, these 
secondary processes presumably do not dominate this study because
\ha/\hb\ approaches the canonical case-B value in quasars which 
we assume to be unobscured (i.e. high $R$ and flat \aopt).

\subsection{Equivalent Widths and Reddening}
\label{sec:aoptew}

\begin{figure}
\centerline{\psfig{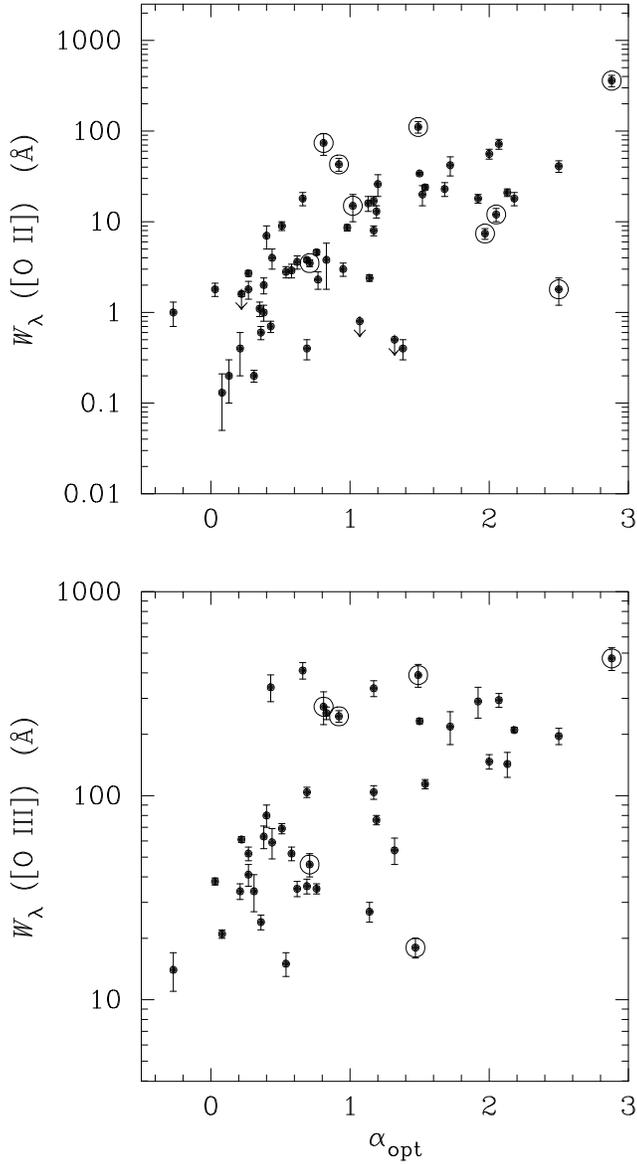} }
\caption[Dependence of Narrow-Line Equivalent Widths on Spectral Index]
{Dependence of narrow-line, \oii\ ({\it top\/}) and
\oiii\ ({\it bottom\/}) equivalent widths, \ew, on optical 
spectral index, \aopt, for the MQS. 
CSS quasars are circled.
  }
\label{nlewaopt}
\end{figure}

To illustrate the link between reddening and narrow-line \ew, 
Figure \ref{nlewaopt} shows the 
equivalent widths of narrow \oii\,\gl3727 and \oiii\,\gl\gl4959,5007 
plotted against the optical spectral index, \aopt. Both \ew\,(\oii) 
and \ew\,(\oiii) correlate remarkably strongly with \aopt, with 
correlation probabilities $>99.99$\% and slopes $0.66\pm0.10$ for \oii\ 
and $0.36\pm0.08$ for \oiii. Such a relationship confirms the view that
the range in \ew\ is due predominantly to the joint weakening and 
steepening of the optical continuum as a consequence of extinction. 
As expected, the Balmer decrements also correlate with the narrow 
\oiii\ and \oii\ equivalent widths.
 
\begin{figure}
\centerline{\psfig{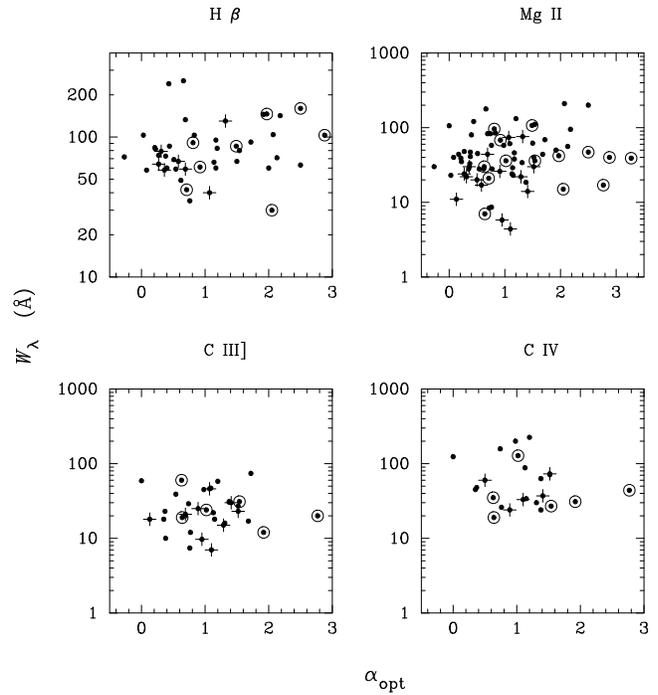} }
\caption[Dependence of Broad-Line Equivalent Widths on Spectral Index]
{Dependence of broad-line, \hb, \mgii, \ciii\ and \civ,
equivalent widths, \ew, on optical 
spectral index, \aopt, for the MQS. 
Error bars are omitted for clarity.
CSS quasars are circled; quasars with $R>1$ have crosses superposed.
 }
\label{blewaopt}
\end{figure}

Figure \ref{blewaopt} shows broad-line \ew\ as a
function of \aopt. No strong correlations with \aopt\ are observed for any of 
the lines, \hb, \mgii, \ciii\ or \civ.

Assuming the optical spectral slope is a statistical 
reddening indicator, the results in Figure 
\ref{blewaopt} are consistent with reddening of both the broad lines 
and continuum by a similar amount in each object. The narrow lines, 
in contrast, must be reddened less than the continuum because their 
equivalent widths are strongly dependent on \aopt. In other words, 
in the reddest objects, both the continuum and the broad lines are 
weak (i.e. broad \ew\ constant), but the narrow lines remain strong (i.e. 
\ew\ large). From these arguments, the material 
causing the extinction must be physically associated with the quasar
itself, and located at radii {\it outside\/} both 
the BLR and optical continuum source but {\it inside\/} the 
narrow-line region (NLR).

\subsection{Testing the Reddening Hypothesis}

Next, the hypothesis that dust extinction can account for
{\it all\/} the observed anisotropy of the optical continuum, as inferred 
from the $R$-dependence of the narrow oxygen-line equivalent widths, 
is tested semi-quantitatively. This comparison assumes
that the range in \ew\ of a given forbidden line
is due solely to changes in the strength of the underlying continuum. 
Also, the Balmer lines and continuum are assumed to suffer the same 
extinction. An approximate $\lambda^{-1}$ reddening law is used (as 
in Osterbrock 1989). 

Increasing the Balmer decrement from \ha/\hb$=2.85$ to 10 
(i.e.~an equivalent $A_{\rm V}=3.7$) in low-$R$ quasars 
(Figure \ref{baldecr}) predicts about 5~mag of 
continuum extinction at 3727\AA. This is consistent 
with the observed range of \oii\  equivalent widths in the MQS 
(Figure \ref{rewo2}). At the wavelength of \oiii, 4~mag extinction is 
expected at low $R$ for the same $A_{\rm V}$ {\it but\/} 
\ew\,(\oiii) varies less than this (approximately 2.5~mag). The 
missing 1--2~mag can be accounted for if the \oiii\ line luminosity  
(but not \oii) increases by this amount over the range $0.01<R<10$. Reddening
a continuum with \aopt$=0.5$ with $A_{\rm V}$ up to 3.7, the predicted 
range of \aopt\ is 0.5-4, i.e. slightly steeper than 
observed (maximum \aopt$\approx 3$).
This may indicate that either the intrinsic spectra are 
flatter than  \aopt$=0.5$ (eg.~Binette, Fosbury \& Parker 1993) or the 
broad lines are reddened more than the continuum by dust
internal to the Balmer-line-emitting clouds (e.g.~Wills et al. 1993).

If a smaller amount of reddening is postulated, enough to explain only the
\ew\,(\oiii) trend with $R$, a larger discrepancy must be
explained in \ew\,(\oii). As discussed above, such a difference could arise 
from ionisation differences leading to weaker \oii\ luminosities in
core-dominated quasars. Also, a range of Balmer decrements smaller 
than observed would be predicted, i.e. \ha/\hb$\approx 3$--6.

\subsection{A New Perspective}
 
\begin{figure}
\centerline{\psfig{file=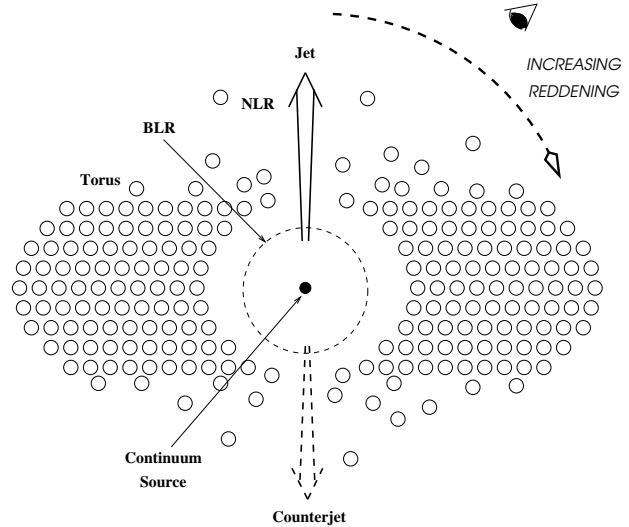,bbllx=1pt,bblly=1pt,bburx=
410pt,bbury=360pt,height=7cm} }
\caption[Torus]
{Cartoon of a reddening geometry for radio-loud AGN; not to scale.
At small viewing angles to the jet, the nuclear regions are relatively
unobscured and a core-dominated quasar is seen.
At larger viewing angles to the radio jet, more
dust is intercepted, reddening the broad lines and continuum in 
lobe-dominated quasars. Beyond some angle, 
the dust becomes opaque and the BLR is obscured.
Such an object would be classified as a radio galaxy.
}
\label{dust}
\end{figure}

The most plausible dust geometry consistent with the reddening trends
discussed above is a torus with its axis aligned with the radio jets 
(see Figure \ref{dust}).
Traditionally, an  {\it opaque\/} torus with a sharp opening 
angle has been proposed to obscure the BLR in narrow-line AGN 
(Krolik \& Begelman 1986).
In contrast, the pronounced reddening observed in lobe-dominated MQS 
{\it quasars\/} implies that significant quantities of dust must lie 
{\it within\/} the putative torus opening angle. In fact there may not be a
well-defined opening angle at all.

In this new picture, the total opacity of
the dusty material observed to redden the broad lines and
optical continua of MQS quasars 
increases systematically as the quasar 
is viewed at larger angles to the jet axis. In directions 
perpendicular to the jet, the column density must be sufficient 
to obscure the broad lines completely, consistent with the unification
of quasars and radio galaxies. Such a strong variation of column density with 
viewing angle might be achieved either geometrically, or 
with a latitude-dependent dust density.

 To prevent the destruction of the dust grains by thermal spallation 
(Krolik \& Begelman 1988), the dust would presumably have to be in a 
clumpy distribution. Therefore, at low opacities, the reddening 
might be expected to be `quantised' with some fraction of quasars at 
any given viewing angle being affected strongly. In this case, the
spectral slopes and Balmer decrements of individual quasars might exhibit 
large scatter with $R$, as in Figure \ref{raopt}, but their average would 
be more representative, consistent with the smooth trends
seen in composite MQS spectra (Baker \& Hunstead 1995). Scatter would
also be expected in the narrow-line \ew--$R$ plots.

The dusty material inferred to be present at high latitudes to the 
plane of the proposed torus may be related to the
scattering clouds along the jet axis which reflect nuclear light 
towards the observer in some AGN (Antonucci \& Miller 1985; Goodrich
1990; di Serego Alighieri et al. 1996). 
Both dust and hot electrons have been proposed as the scattering medium.
Indeed, substantial amounts of reddening would be expected to result 
in dichroic polarisation of the quasar light. Lobe-dominated quasars may
also show polarised broad-line emission, similar to that observed for 
radio galaxies. 
Also, the distribution of the dust in the torus region will have
implications for modelling the infrared emission from quasars (eg.~Pier 
\& Krolik 1992; Efstathiou \& Rowan-Robinson 1995). The observed
anisotropy of the infrared emission (Heckman et al. 1994) may imply
that the dust remains optically thick down to long wavelengths.
At higher energies, reddened quasars might show more UV and/or 
X-ray absorption than their core-dominated counterparts, providing a
valuable probe of the inner regions of the torus. Optical absorption
lines might also be more common in lobe-dominated quasars as a consequence.

Although red quasars have been noted in the literature before
(Smith \& Spinrad 1980; Mathur 1994), their relationship to other quasars
and the origin of their steep continua has remained unclear.
The extinction of the BLR and continuum in quasars viewed 
at large angles to the radio jet is consistent with the unification
by obscuration of radio sources, implying a smooth transition from quasars
to radio galaxies.
An analogous reddening geometry has been proposed for Seyfert galaxies, 
such that extinction is least in Seyferts observed with the minor axis of the
host galaxy oriented close to the line of sight
(de Zotti \& Gaskell 1985).
The aspect-dependent extinction model is also in agreement with 
the large extinctions ($A_{\rm V} \sim 30$) inferred towards narrow-line AGN
(Simpson, Ward \& Wilson 1995)
and recent discoveries of broad Balmer emission lines in the infrared, 
and the subsequent re-classification, of a number of radio galaxies 
(e.g.~Lacy et al. 1995).
In this way, dust would also affect the relative numbers
of classified broad and narrow-line objects, perhaps relieving some
problems with simple interpretation of the radio-loud unified scheme
(Singal 1993; Laing 1994; Kapahi et al. 1995). 
Preferential reddening of lobe-dominated quasars will also introduce yet
another orientation-dependent sample selection bias on the basis of
optical colour, such that faint, red low-$R$ objects are more likely to 
be missed on blue photographic plates. 
It is also possible that the effects of dust extinction 
are more prominent in faint quasar samples, such as the MQS, than in 
brighter samples, prompting the further investigation of the luminosity 
dependence of reddening.

\subsection{Broad Lines}
 
In general, \ew\  is not a strong function of $R$ for the broad lines,
suggesting the lines themselves suffer extinction.
However, there is a tendency for broad 
\ew\ to be weaker in core-dominated quasars; this is not a 
selection effect. The decrease of \ew\ with $R$ is clearest for the 
low-ionisation line \mgii. This trend suggests that these broad
lines are either emitted more isotropically than the continuum, or 
suffer less reddening. Perhaps dust surviving 
at very small radii within the BLR obscures the continuum more than the
broad-line clouds. If dust is distributed throughout the BLR, 
then \ew\ of low-ionisation or semi-forbidden lines in particular might
be expected to show some correlation with $R$, or \aopt: excluding CSS
quasars on  Figure \ref{blewaopt}, the reddest quasars also have 
the highest \ew\,(\mgii).

The roughly constant \ew\ of the Balmer recombination lines, notably \hb,  
must imply that
even if a relativistically-beamed component is present in luminous,
high-$R$ quasars, it does not dominate the photoionising
continuum we see. 
Aspect-dependent reddening would explain naturally the 
constancy of \ew\,(\hb) with $R$, albeit fortuitously. 
An alternative picture is that the optical
continuum originates in the same region as \hb\ (Binette et al. 1993).

The redshift/luminosity dependence of \ew\,(\mgii) is quite 
striking; the line is observable over the largest range of 
redshifts in the MQS. This dependence may  indicate the presence
of either a stronger continuum or a higher ionisation parameter
in high luminosity quasars. Trends with redshift are not seen 
for the narrow \oii\ and \oiii\ lines and the range in redshift 
is too small to judge this for \ciii\ and \civ.


\section{Conclusions}

(1) The analysis of optical spectra for the MQS presented in this paper 
provides strong evidence that the optical continuum emission 
in radio quasars is highly aspect dependent, 
enhanced in core-dominated quasars by 3--5\,mag.

(2) The bulk of the aspect dependence in lobe-dominated MQS quasars, 
both of the optical continuum and the broad emission lines, arises from
anisotropic extinction.
Reddening is favoured from the $R$-dependence of the optical continuum
slope and Balmer decrement; it is favoured similarly by the wavelength 
dependence of the continuum changes, varying {\it more\/} with 
viewing angle in the blue. Furthermore, the values of \ha/\hb\ in unobscured 
quasars (i.e. high $R$, flat \aopt) support the `case-B' 
recombination picture.

(3) Extinction affects the continuum and broad lines more than the
narrow lines, suggesting that the dust distribution responsible 
is  physically associated with the active galaxy and lies 
at distances between the BLR and NLR.
A dusty torus is a plausible candidate. However, substantial numbers 
of dust clouds  must lie within the torus opening angle, contributing to
the average column density that increases with viewing 
angle away from the radio-jet axis. 

(4) The ratio of \oii/\oiii\ is also $R$-dependent, consistent with
anisotropic obscuration of \oiii\ at small radii and \oii\ being more
isotropic than \oiii\ and arising at larger distances from the nucleus.

(5) Aspect-dependent reddening in quasars implies
that magnitude-limited samples must be affected strongly by 
orientation-dependent selection biases, affecting not only 
optical brightness but also colour and broad-line properties.


\section*{Acknowledgments} 

Special thanks to Richard Hunstead, Vijay Kapahi, C.R. Subrahmanya and
Pat McCarthy for extensive work on the identification and follow-up 
of the MQS. Thanks also to the referee, Neal Jackson, plus
Richard Saunders, Ian Browne, Peter Barthel, 
Steve Rawlings and Chris Simpson for 
valuable discussions and Raylee Stathakis and
Elaine Sadler for AAT Service observations. The 
AAT observatory staff are also thanked. JCB acknowledges a
graduate scholarship from the Research Centre for Theoretical
Astrophysics, University of Sydney, for part of this work.


\bsp
\end{document}